\documentclass[pra,twocolumn,superscriptaddress]{revtex4}
\usepackage[koi8-r]{inputenc}
\usepackage[english]{babel}
\usepackage{amsmath}
\usepackage{amssymb}
\usepackage{epsfig}
\begin{document}

\title{Quasi-stable quantum vortex knots and links in anisotropic harmonically trapped Bose-Einstein condensates}
\author{Christopher Ticknor}
\email{cticknor@lanl.gov}
\affiliation{Theoretical Division, Los Alamos National Laboratory, Los Alamos, New Mexico 87545, USA}
\author{Victor P. Ruban}
\email{ruban@itp.ac.ru}
\affiliation{Landau Institute for Theoretical Physics RAS,
Chernogolovka, Moscow region, 142432 Russia} 
\author{P.~G. Kevrekidis}
\email{kevrekid5@gmail.com}
\affiliation{Department of Mathematics and Statistics, University of Massachusetts,
Amherst, Massachusetts 01003-4515 USA} 
\date{\today}

\begin{abstract}
Long-time existence of topologically nontrivial configurations of quantum vortices 
in the form of torus knots and links in trapped Bose-Einstein condensates  
is demonstrated numerically within the three-dimensional Gross-Pitaevskii equation
with  external anisotropic parabolic potential. We find out parametric domains near the 
trap anisotropy -- axial over planar frequency trapping ratio $\lambda\approx 1.5-1.6$ 
where the lifetime of such quasi-stationary rotating vortex structures is many hundreds 
of typical rotation times. This suggests the potential experimental observability of
the structures. We quantify the relevant lifetimes as a function of the model parameters 
(e.g. $\lambda$) and initial condition parameters of the knot profile.
\end{abstract}
\maketitle

\section{Introduction}
Topological structures bearing vorticity have been long recognized as objects of high 
interest in hydrodynamics, optics, 
and condensed-matter physics \cite{Saffman,Pismen, Donnelly}. 
Within the particular theme of
atomic gases in the realm of 
Bose-Einstein condensates~\cite{pita1,pita2,ourbook}, a pristine
setting has been identified for the
exploration of the properties of 
such structures. More specifically,
the static and dynamical properties of quantized vortices have played a crucial role
in a wide range of associated theoretical, numerical and experimental studies;
as only a small ensemble of relevant examples, we mention the 
reviews~\cite{FS2001,F2009,PGK2004,Kom2007,ParkerBar,andAJP}. 

A focal theme of interest within
this nexus of topological charge, nonlinearity and spatial confinement
has been the study of vortex rings and simple filaments
~\cite{SF2000,R2001,AR2001,GP2001,AR2002,RBD2002,AD2003,AD2004,D2005,Kelvin_waves,ring_instability,v-2015,BWTCFCK2015,R2017-2,R2017-3,reconn-2017,top-2017,WBTCFCK2017,TWK2018}
whose interaction dynamics and even leapfrogging~\cite{promentbar,talley,R2018PoF} 
have been considered. An even more demanding
3D territory that has been less explored (especially so experimentally) has been that
of vortex knot structures.
These have been examined mainly for a uniform density background; see \cite{RSB999,MABR2010,POB2012,KI2013,POB2014,LMB016,KKI2016,R2018-2,R2018-3},
and references therein. Also, no experimental technique for producing knots and links 
in Bose-Einstein condensates has been developed, in the exception of the remarkable 
synthetic structures produced in spinor Bose-Einstein condensates~\cite{hall1,hall2}.

Very recently in Ref.~\cite{R2018-1},
based on the hydrodynamic approximation (with potential perturbations neglected), 
simple vortex knots were theoretically considered in trapped axisymmetric condensates 
characterized by an equilibrium density profile $\rho(z,r)$. In particular,
stability of torus vortex knots under suitable conditions was predicted.
Its preliminary numerical verification was undertaken very recently by one of 
the present authors
(V.P.R. \cite{R2018-4}) within the Gross-Pitaevskii (GP) equation, 
with the latter representing a suitable
three-dimensional (3D) model for a rarefied Bose gas at zero temperature. 
For a few sets of system parameters, long lifetimes for torus vortex knots, 
unknots, and links were indeed observed.
On the other hand, it is important 
to highlight that the earlier systematic
work of~\cite{KKI2016} (involving thousands
of relevant simulations) predicted 
instability of all the examined types
of knots in the homogeneous condensate
cases considered therein.

In light of the above results, there
is an important open question remaining.
Can knot (or link) structures become
dynamically robust in the presence of
trapping~? Here, we examine this question
in the context of variation of model
parameters and initial condition parameters.
The former are represented by the parametric exploration as a result,
e.g., of the trap anisotropy, while the latter are induced by the variation  of
the vortical pattern initial locations.
Given the generic rotation exhibited by
knot patterns, we do not seek these
as exact stationary solutions. Rather,
we consider a large range of dynamical
simulations where a perturbed initial
configuration is evolved and the outcome
of the evolution is assessed, attempting in this way to offer a systematic 
view of the knot lifetime problem.
The relevant extensive numerical simulations  suggested, among other things,
a definite optimization (maximization) 
of the vortex knot lifetimes for values of
$\lambda$ (the axial vs. planar trapping strength) around 1.5-1.6. 
They also revealed that in the trapped
setting, distinct destabilization
pathways may arise for the knots.
In particular, they may not only
``untie'' as they do in the homogeneous
setting, but rather portions of the know
may exit the region (confinement induced) 
of non-vanishing density, thus destroying the structure.
We now turn to the relevant theoretical setup and the
corresponding detailed numerical findings.

\section{Theoretical Setup and Numerical Method}
The 3D Gross-Pitaevskii equation in trap units~\cite{note} takes the form 
\begin{equation}
i\psi_t=\Big[-\frac{1}{2}\Delta
+\frac{1}{2}(r^2+\lambda^2 z^2)+g|\psi|^2-\mu\Big]\psi,
\label{GP}
\end{equation}
where $r^2=x^2+y^2$.
The principal parameters here are the trap anisotropy $\lambda$ 
(the ratio between axial and planar trapping strengths)
and the interaction strength: $g=4 \pi N a/l_r$ 
(however, appropriate re-scaling of $\psi$ is able to give $g=1$).
Here $a$ is the s-wave scattering 
length, $l_r$ is the oscillator length: $\sqrt{\hbar/m\omega_r}$,
$m$ is the atomic mass, and $\omega_r$ is the planar trap frequency.
The chemical potential $\mu$ is assumed sufficiently large 
(here we typically use $\mu\sim 30\hbar\omega$
unless indicated otherwise), 
in order to ensure the hydrodynamic 
--referred to also as  Thomas-Fermi-- regime. 
As a result of this regime, the equilibrium condensate density
can be well described by the expression:
$\rho(z,r)\propto [\mu-(r^2+\lambda^2 z^2)/2].$
Thus, the ellipsoid $r^2+\lambda^2z^2=2\mu$, with transverse size $R_\perp=\sqrt{2\mu}$,
is an effective boundary of the condensate at equilibrium, i.e., its density vanishes
outside of this ellipsoid. Following Ref.~\cite{R2018-1},
in the deep Thomas-Fermi limit $\mu\gg 1$ quasi-stationary vortex torus knots and links 
$T_{p,q}$ are possible. We have knots when $p$ and $q$ are co-prime integers 
(including the case $p=1$ and/or $q=1$ of trivial knots that can be unfolded 
to a ring -- ``unknots''), and we have links when $p=np'$, $q=nq'$, with $n\geq 2$ 
($n$ knots or rings that are linked together). For example, the well-known trefoil knot 
is $T_{2,3}$, while the Hopf link is $T_{2,2}$ \cite{hopflink}.
All such structures were theoretically found in Ref.~\cite{R2018-1} to have 
equilibrium toroidal radius $R_*(\mu)=\sqrt{2\mu/3}$, 
and the healing length at that radius is $\xi_*=\sqrt{3/(2\mu)}$. 

The initial (condition for the) position 
of the vortex core in our studies is assumed 
to be a distorted torus knot
(links are constructed in a similar manner)
\begin{equation}
r(\varphi)+iz(\varphi) =r_0+r_1e^{iw\varphi} +\sum_m A_m e^{i(m\varphi/p+\gamma_m)},
\label{distorted_knot}
\end{equation}
where $w=q/p$ is the winding number, $r_0$ ($r_1$) is the toroidal (poloidal) radius,
$A_m$ and $\gamma_m$ are real amplitudes and phases of perturbations. 
The latter are needed to break the symmetry of the knot and thus introduce ``seeds''
for the development of possible instabilities.
Two variants of vortex shape are studied in our numerical experiments:

(S1) we use $r_0$=4.0, $r_1$=0.7 and for a single $m$ we take $A_m=r_1/20$ and 
$\gamma_m=0$, while all the remaining amplitudes are set to zero;

(S2) the sum in Eq.~(\ref{distorted_knot}) is taken over a finite range 
$(q-10)\leq m\leq (q+10)$, with all-equal $A_m$'s from the set 
$\{0.001,0.005,0.010\}R_*$, and with quasi-random $\gamma_m$'s uniformly distributed
on interval $[0:2\pi)$.

In case S2 a typical value of the sum is about $5A_m$ 
which should be compared to $r_1\sim 0.20 R_*$.
So,  $A_m=0.001R_*$ gives a nearly perfect torus knot, 
while $A_m=0.01R_*$ results in significant distortions.

Now to construct the full 3D initial condition we need to specify all 
of the vortex cores in the $(r,z)$ plane
for a given $\phi$ where there are $p$ vortices, here $r>0$.
We are able to construct the phase of the wave function with
the superposition of the phase from each vortex:
\begin{eqnarray}
\Psi(\phi,r,z)/\sqrt{|\rho|}=\Pi_j^p \psi_{2D}(r-r_j,z-z_j) \label{phase}
\end{eqnarray}
where $r_j$, $z_j$ is the position of the $j$-th vortex core and 
$\psi_{2D}(r,z)= e^{i \theta}$ with $\theta=\mbox{atan2}(r,z)$ 
where $r$ and $z$ are the distance to a vortex.  Thus, the total phase is simply 
the product of all the vortex core phases.

Additionally, we use a multi-step algorithm to find the ``ground'' state of the 
relevant system. More specifically:

(i) We imprint the phase of the ground state as found from Eq.~(\ref{phase}).

(ii) We temporarily introduce an additional pinning potential defined by the sum 
$V(\phi,r,z)=U\sum_j e^{-B\Theta_j}$ where $\Theta_j=(z-z_j)^2+(r-r_j)^2$, 
$U$ and $B$ are suitable coefficients. We concentrate mainly on the following two choices: 
(V1) a relatively smooth pinning with $U=50,B=15$;
(V2) a sharp pinning with $U=600,B=240$.

(iii) We have a short, but heavily damped 
imaginary time propagation corresponding to a dissipative regime. 
This step allows a relaxation of wave function in the trap that eliminates 
large-scale sound-mode perturbations \cite{R2018-4}. 

It should be noted however that despite the pinning, the vortex core still 
retains some small deviation from the prescribed shape (\ref{distorted_knot}) 
during the dissipative stage. Mainly it is a small increase of $r_1$. 
But with our $U$ and $B$, the deviation is less than vortex core width.
Another important point is that the relatively smooth pinning (V1) potential 
results in a ``fat'' vortex
core at the end of stage (iii). As conservative evolution starts, the core returns 
quickly to its normal width, thus producing some short-scale non-stationary 
ripples on the density background. The ripples act then as additional perturbations 
and reduce the vortex lifetime comparatively to more clean backgrounds corresponding 
to the sharper pinning potential (V2). However, further sharpening of the pinning
potential is not efficient as it is
unable to trap the vortex.

The time propagation of Eq.(\ref{GP}) takes place with a third-order
operator splitting Fourier spectral method, with time time steps of $5\times 10^{-4}$,
with a numerical grid of 256${}^3$, and with a spacing of 0.07$l_r$. 
This method preserves energy at the 8th decimal place for all simulations.

Having provided the setup of our numerical experiments, we now
turn to a summary of our extensive numerical investigations.

\section{Results}

The behavior of a knot in free space has been studied, in particular, 
in Refs. \cite{POB2012,KKI2016}.
The basic motion of a trefoil is that the knot rolls over in a regular motion. 
Eventually the knot will untie as perturbations grow and the regular rolling 
motion ends~\cite{POB2012,KKI2016}. 
We will now examine the behavior of a knot in various geometries, 
starting with $\lambda$ equal to 0, 0.85, and then looking at 1.6 and 1.8.
We present results for trefoil knots with a single-$m$ perturbation S1 
and smooth pinning V1.
\begin{figure}
\includegraphics[width=0.48\columnwidth]{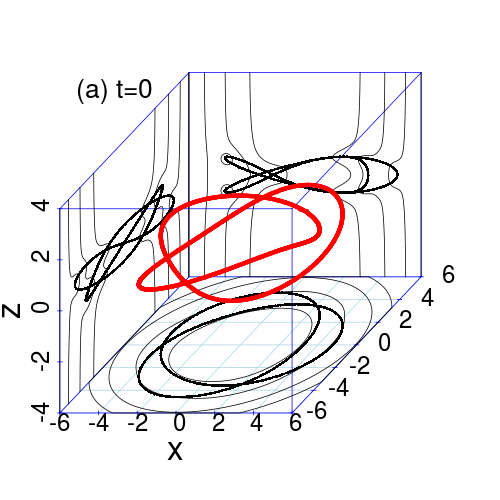}
\includegraphics[width=0.48\columnwidth]{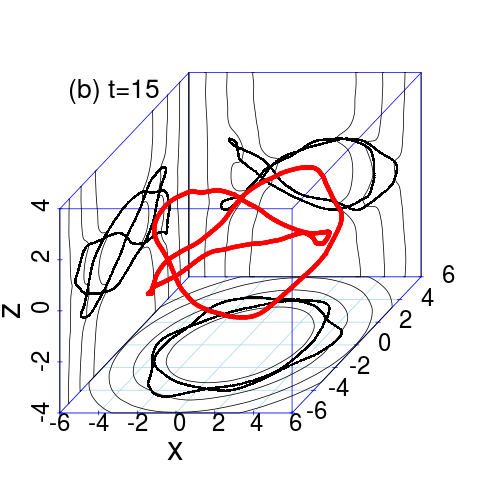}
\includegraphics[width=0.48\columnwidth]{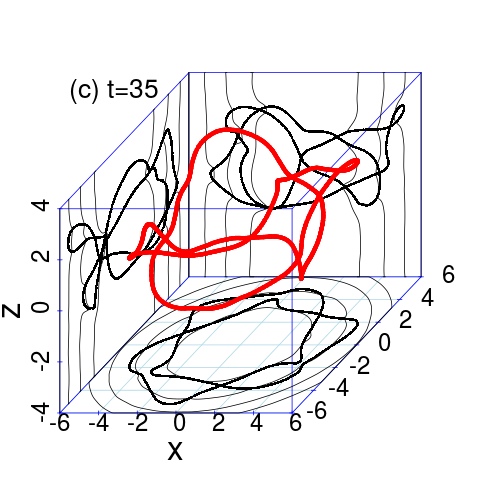}
\includegraphics[width=0.48\columnwidth]{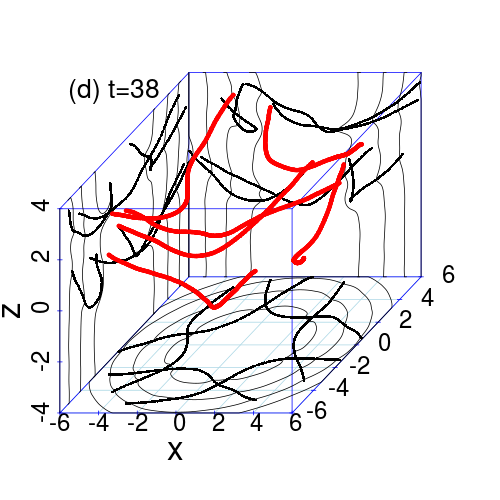}
\caption{
For $\lambda=0$ snapshots are shown 
along the demise of the knot within the BEC.
(a) t=0, the initial condiguration, (b) the knot distorts,
(c) portions of the knot further extend outward, and (d) portions 
of the vortical pattern leave the volume.
The axes are in oscillator units, $\sqrt{\hbar/m\omega_r}$.
}\label{knot0}
\end{figure}

In Figure \ref{knot0} we show the evolution of a $T_{23}$ knot
in a trap with $\lambda=0$. This is just a tube scenario, involving no confinement 
in the z direction.
The red line is the 3D vortex, i.e., it
represents the position of the vortex core. 
The vortex positions are extracted by finding the phase singularity 
on the computational grid \cite{FBD2010}. We further refine these vortex positions via  
method used in Ref. \cite{top-2017}.
Additionally, both the BEC's density (thin black lines) and the extracted cores 
are projected (bold black) onto the back planes: $(x,y)$, $(x,z)$, and $(y,z)$. 
One can discern that early on during
the evolution for this scenario the knot gets distorted due to 
undulations (the so-called Kelvin 
waves~\cite{Kelvin_waves,ourbook}).
As a result, already at times earlier than $40$ in our dimensionless units,
the knot has broken into individual undulating filaments, losing its coherence
as a trefoil structure.

\begin{figure}
\includegraphics[width=0.48\columnwidth]{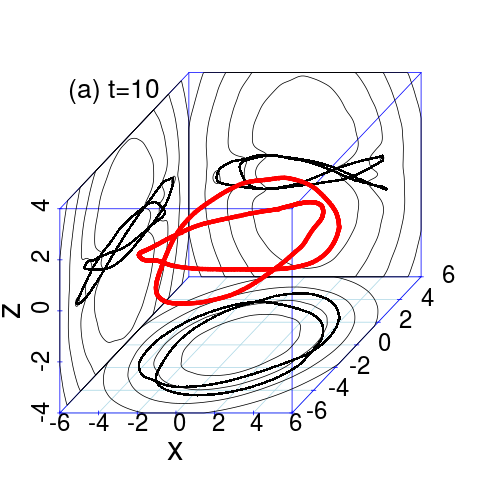}
\includegraphics[width=0.48\columnwidth]{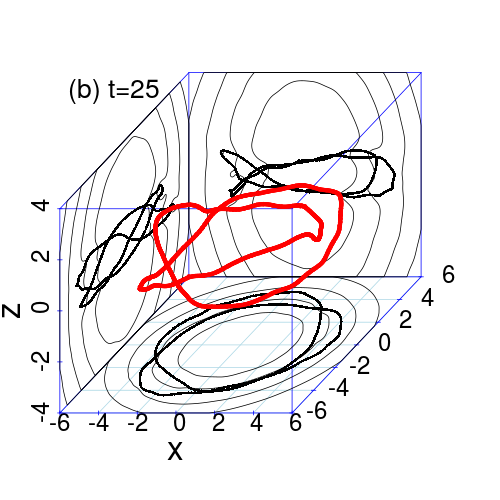}
\includegraphics[width=0.48\columnwidth]{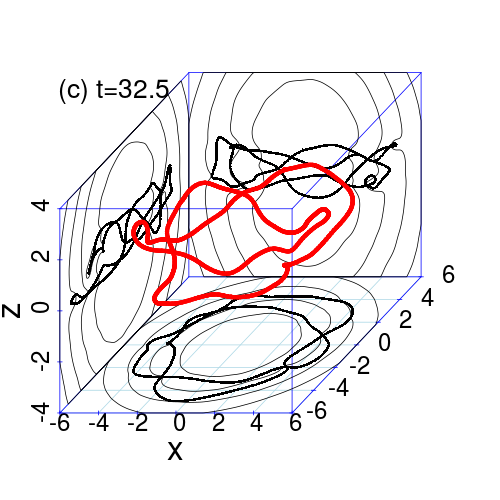}
\includegraphics[width=0.48\columnwidth]{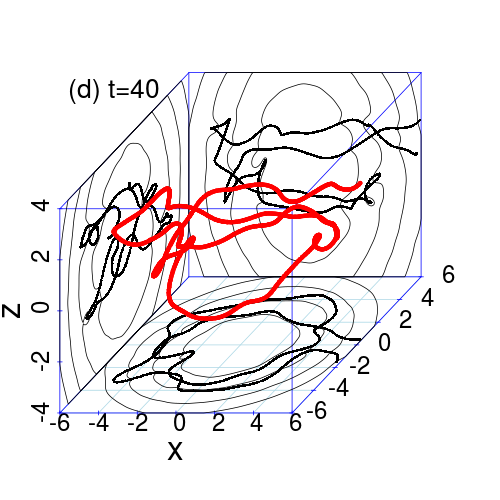}
\caption{
For $\lambda=0.85$ snapshots showing the decay of the vortex
knot.  The evolution already looks very different for $\lambda=0$.
In (a) we see the early form of the knot at t=10;
(b) at t=25 some Kelvin-wave induced
undulations arise.
(c) At t=32.5 the knot has untied, and 
(d) later at t=40 a loop has left the BEC.
The axes are in oscillator units, $\sqrt{\hbar/m\omega_r}$.
}\label{knot85}
\end{figure}

The above unconfined along the
z-direction scenario can be compared/contrasted
with the trapped case along the z-direction.
In Fig.~\ref{knot85} we show a system with $\lambda=0.85$. 
In this case the knot unties, more like the knot's evolution in free
space.
The knot is shown at various stages of its evolution with an $m=4$ 
perturbation. At (a) t=10 we see early form of the knot, while
at t=25 (panel (b)) the knot has started
featuring Kelvin-wave undulations.
(c) At t=32.5 the knot has untied, and 
(d) later at t=40 a loop has left the BEC's volume.
For the $\lambda=0.85$ this is fairly typically behavior.
With lower $m$ (1, 2, 3) perturbations, the knot also tilts 
like a ring \cite{TWK2018} in a trap with $\lambda<1$. But the knot 
still unties in a similar fashion. 

\begin{figure}
\includegraphics[width=0.48\columnwidth]{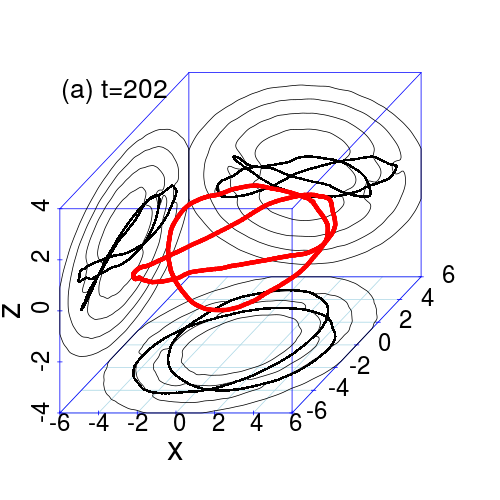}
\includegraphics[width=0.48\columnwidth]{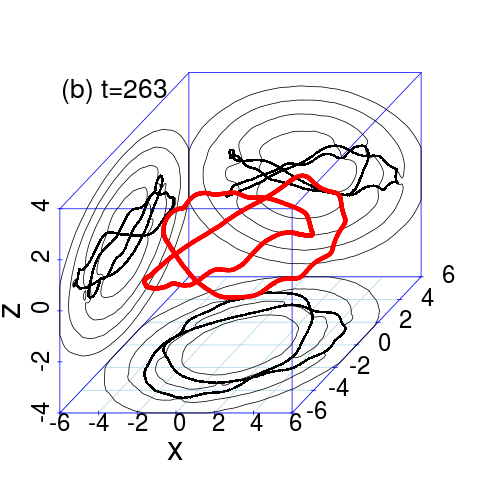}
\includegraphics[width=0.48\columnwidth]{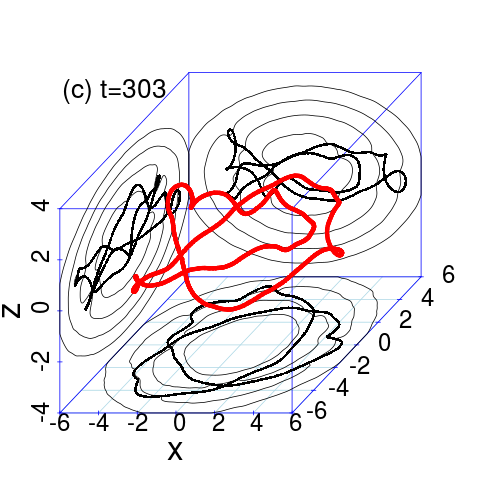}
\includegraphics[width=0.48\columnwidth]{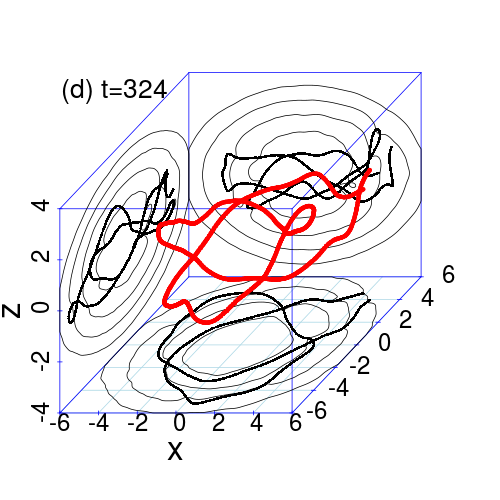}
\caption{
For $\lambda=1.8$ snapshots are shown 
along the demise of the trefoil knot inside
the BEC.
(a) $t=202$ initial distortions appear;
(b) $t=263$ growth of undulations appears.
(c) At $t=303$ further growth of undulations
is shown, while (d) shows the
eventual breakup around at $t=324$ when a portion of the
knot leaves the BEC's volume.
The axes are in oscillator units, $\sqrt{\hbar/m\omega_r}$.
}\label{knot18}
\end{figure}

In a trap, it is important
to highlight that (in addition to 
untying) 
there is another way for the knot to decay:
perturbations can grow so large that a portion of the knot can leave 
the BEC's Thomas-Fermi ellipsoidal
confinement region before the know unties. 
In Figure \ref{knot18} we show the evolution of a $T_{23}$ knot
in a trap with $\lambda=1.8$ with no perturbation, $A_m$=0. 
Here, it can be seen that the evolution
retains the coherence of the trefoil
for times that are about an order of magnitude
longer than $\lambda=0$ and 0.85. 
Already in panel (b) at $t=263$, 
the helical, Kelvin-like
undulations have started forming.
These are more substantially amplified 
at $t=303$ (panel (c)) where the knot further distorts but still has a trefoil structure.
{Finally in (d) at $t=324$ the structure
becomes ``untied'' not by becoming a link, but 
by having a portion leave the volume (i.e.,
the confinement region discussed above).}

\begin{figure}
\includegraphics[width=0.48\columnwidth]{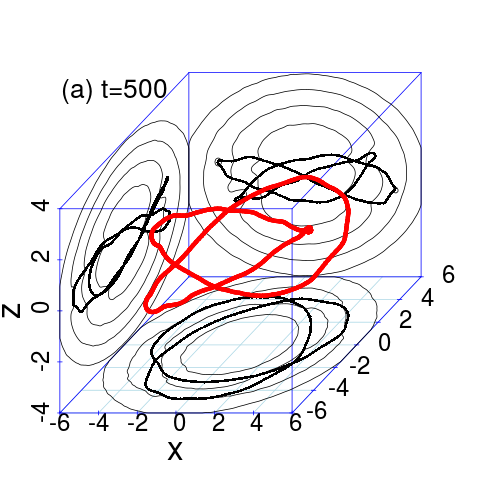}
\includegraphics[width=0.48\columnwidth]{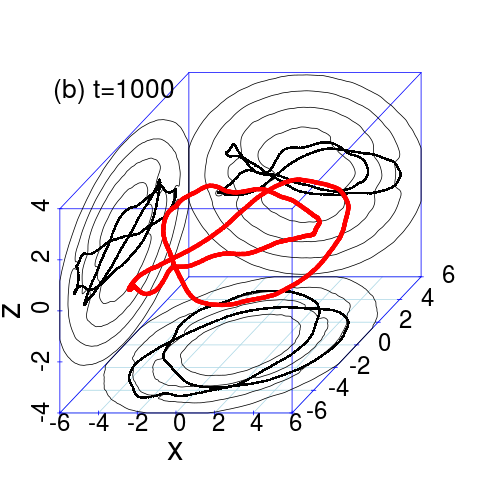}
\includegraphics[width=0.48\columnwidth]{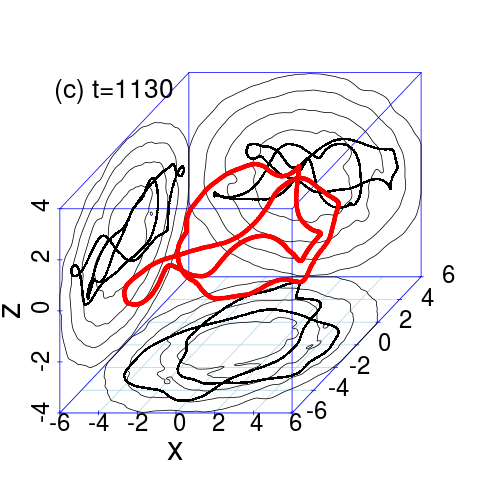}
\includegraphics[width=0.48\columnwidth]{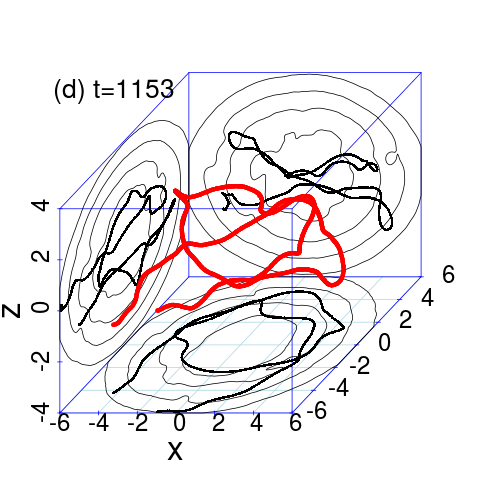}
\caption{
For $\lambda=1.6$ snapshots showing the decay of the vortex
knot. At (a) $t=500$ and (b)$1000$ the emergence of nontrivial
undulations can be observed but 
these remain small.
(c) At $t=1130$ the knot has untied into a link,
while
(d) at $t=1153$ a portion of the link is leaving the volume.
The axes are in oscillator units, $\sqrt{\hbar/m\omega_r}$.
}\label{knot16}
\end{figure}

To illustrate the main point of
our work, namely the dramatic
impact of judiciously chosen anisotropy
on the lifetimes of the vortex
knots, we now turn to a case 
involving $\lambda=1.6$.
In Figure \ref{knot16} we show the evolution of a $T_{2,3}$ knot
in a trap with this $\lambda$. 
{The knot lives over 1100 trap units of time before
it unties. Just after the knot unties, it 
is shown in (c), and then the knot evolves and eventually a portion of the structure
leaves the volume in (d).}
Remarkably, under similar initialization
as in the cases considered above,
we observe a lifetime about 4 times larger
than in Fig.~\ref{knot18} and nearly
30 times longer than for the case of
Fig.~\ref{knot0}. It is then clear
that a suitable tuning of the anisotropy
can endow a knot structure with very long
life times, conceivably enabling its
observability in already available,
state-of-the-art experimental BEC setups.

\begin{figure}
\includegraphics[width=0.98\columnwidth]{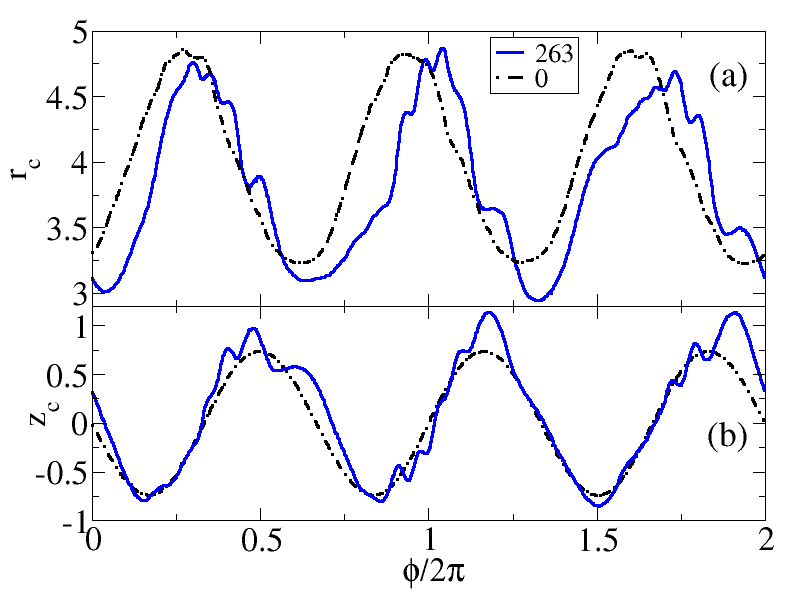}
\includegraphics[width=0.98\columnwidth]{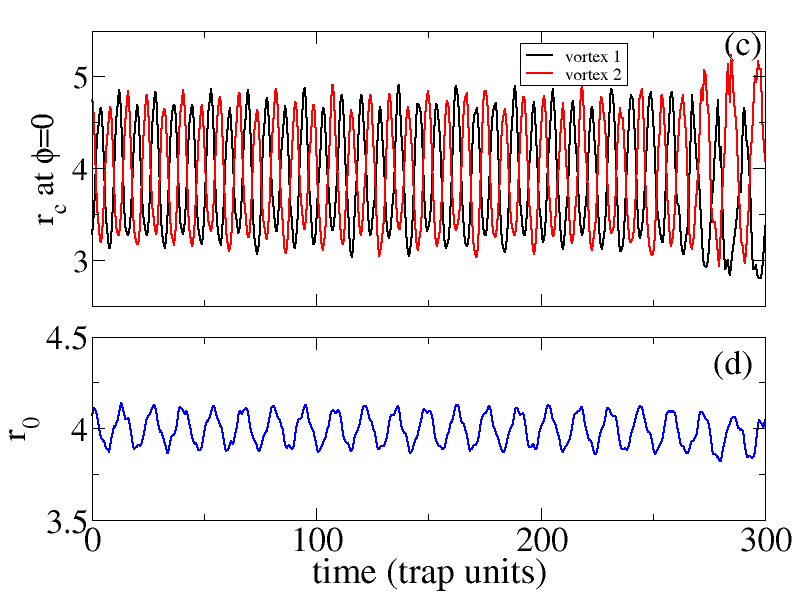}
\caption{For $\lambda=1.8$ (a) $r_c$ and (b) $z_c$ as a function of $\phi$
are shown at different times 0 (black dash-dot) and 296 (blue). 
Also see Fig. \ref{knot18}(a) where the same data is displayed in 3D space.
From these data we extract $r_0$, $r_1$, and $z$.
(c) The radial position of each vortex at $\phi=0$ as a function of time.  
(d) The average $r_0$ as a function of time.
The y axes are in oscillator units, $\sqrt{\hbar/m\omega_r}$.
}\label{phi}
\end{figure}

To measure the lifetimes of the knot
structures, we analyze the extracted core positions 
within the (approximate Thomas-Fermi) region $0.9 R_\perp$. 
(The $0.9$ prefactor is used to avoid ghost vortices from disrupting the knot 
analysis \cite{top-2017,FBD2010}.) 
Then, we order the core positions so that they are a continuous function
in $\varphi$ spanning the interval 
from 0 to $2 p \pi$; see Fig. \ref{phi} (a) for $r$ (b) $z$ of the extracted vortex
core positions.

When the separation of two points anywhere along the knot exceeds a cutoff distance 
(larger than the grid spacing) the knot is considered to be broken.
This works also for reconnection events as a knot unties. 
To further the analysis we can extract the toroidal $r_0(t)$; 
poloidal $r_1(t)$; and $z$, labeled $z_0(t)$  positions of the vortex in the knot. 
Using these quantities, we define $\langle z\rangle $ by taking the average of the $z_0$ 
coordinate over the entire knot: $\langle z \rangle=1/N_c\sum_i^{N_c} z_i$
where $N_c$ is the number of vortex core positions found on our grid;
$r_0=\langle r \rangle$ is found in the same fashion. 
To obtain $r_1$ at a given time we take 
$r_1^2=1/N_c\sum_i^{N_c} (z_i-\langle z \rangle)^2+(r_i-\langle r \rangle)^2$. 
In Fig. \ref{phi}(c) we show the radial position of the two vortex cores in a knot 
for $\phi=0$ as a function of time. 
The position of one core is red while the the other is black.
We can see their regular motion as they tumble over each other.
We can see $r_0$ is the average radial position of the knot and 
and $r_1$ as the distance the two vortices traverse in their 
rotation about each other. 
In Fig. \ref{phi}(d) we show the extracted $r_0$,
note the vertical scale difference of (c) and (d).

\begin{figure}
\includegraphics[width=0.99\columnwidth]{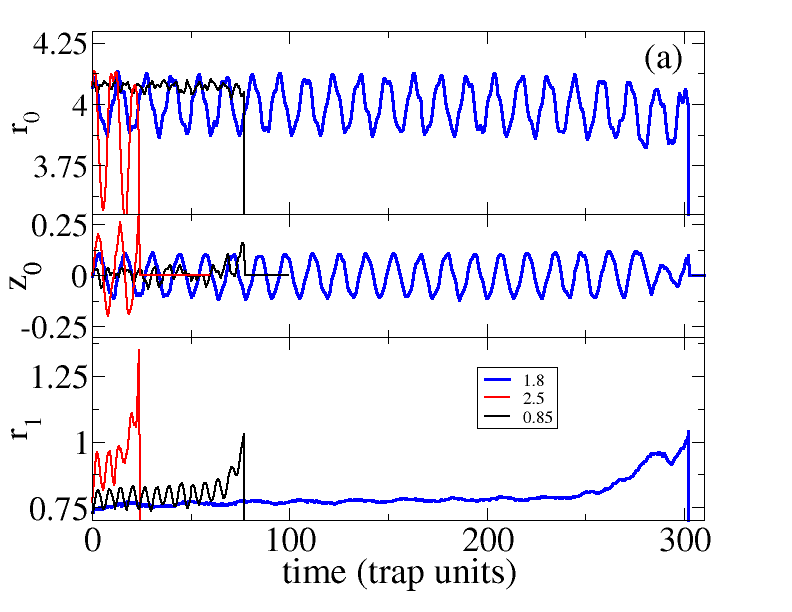}
\includegraphics[width=0.99\columnwidth]{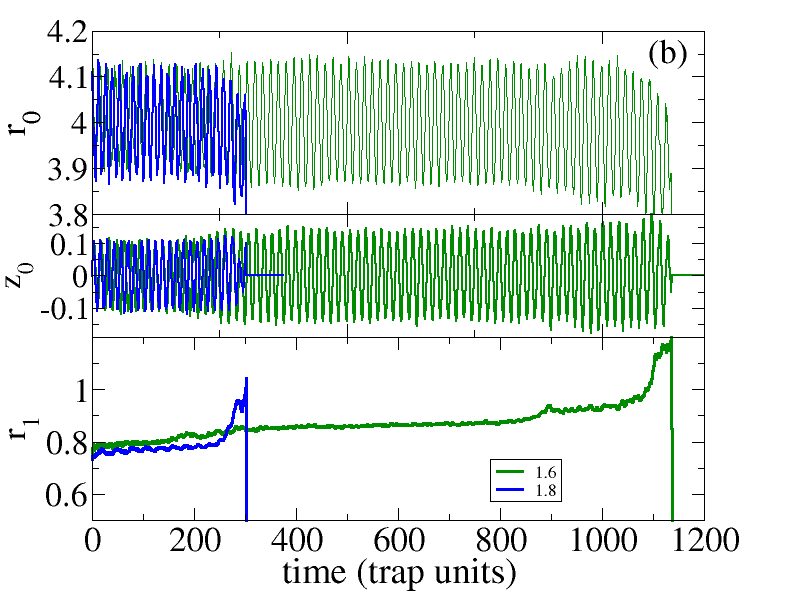}
\caption{
The extracted effective coordinates  $r_0$ (top panel), $z_0$ (middle panel), 
and $r_1$ (b) are shown for different $\lambda$'s.
In (a) $\lambda$=0.85 (black), 1.8 (blue), and 2.5 (red) are shown.
In (b) $\lambda$=1.6 (green), 1.8 (blue) are shown.
It is important to note the disparity in
the time scales of the breakup of the different anisotropy knot structures.
The same initial knot and chemical potential
were used. 
When the knots are no longer complete (untie), the curves are set to zero.
The y axes are in oscillator units, $\sqrt{\hbar/m\omega_r}$.
}\label{timerz}
\end{figure}

We contrast the typical evolution of $r_0$, $r_1$ and $z_0$ for different $\lambda$'s 
in Fig. \ref{timerz}.
In (a) and (b) we show the extracted 
$r_0$ (top panel), $z_0$ (middle panel)  and $r_1$ (bottom panel) for many different geometries. 
In (a) we show the shorter evolution of  $\lambda$= 0.85 (red), 2.5 (lack), and 1.8 (blue). 
Then in (b) we compare the long evolution of  $\lambda$= 1.6 (green) and 1.8 (blue).
The main observation here, corroborating
the snapshots presented earlier
in Figs.~\ref{knot85}, \ref{knot18}, \ref{knot16}  is that we have a huge variation in the 
lifetime of trapped knots.  
In particular, while for prolate or
highly oblate condensates the knots are highly 
unstable [see the black and red
curves for $\lambda=0.85$ and $\lambda=2.5$
in (a)], it is possible
to expand their lifetimes by over 
a factor of 10, by judiciously tuning
the anisotropy towards somewhat oblate condensates, most notably
in the case of $\lambda=1.6$, green curve in (b). 
We observe nearly 100 rotations of the knot in the $\lambda=1.6$ case. 
Interestingly (but also perhaps somewhat intuitively)
this also reflects the corresponding
earlier observations for the stability
in the case of vortex rings which can be thought of 
partial constituents of vortex knots. See, e.g., 
the theoretical analysis of~\cite{ring_instability},
as well as the recent numerical confirmation of~\cite{TWK2018}.

The oscillations in these quantities ($r_0$, $r_1$, and $z_0$) are 
related to the excitations of the knot, but also correspond to 
the knot completing a rotation, see 
Fig.~\ref{phi}.
Additionally, it is relevant to note
within Fig.~\ref{timerz} that there
is a certain ``universality'' in the
way that the knot structure manifests
its demise in these diagnostics. 
For most cases we have looked at (except the $\lambda$=0), 
the knot breaks in a fashion where the 
poloidal (effective) coordinate $r_1$
seems to diverge as a portion of the knot leaves the volume or unties.

\begin{figure}
\begin{center}
\epsfig{file=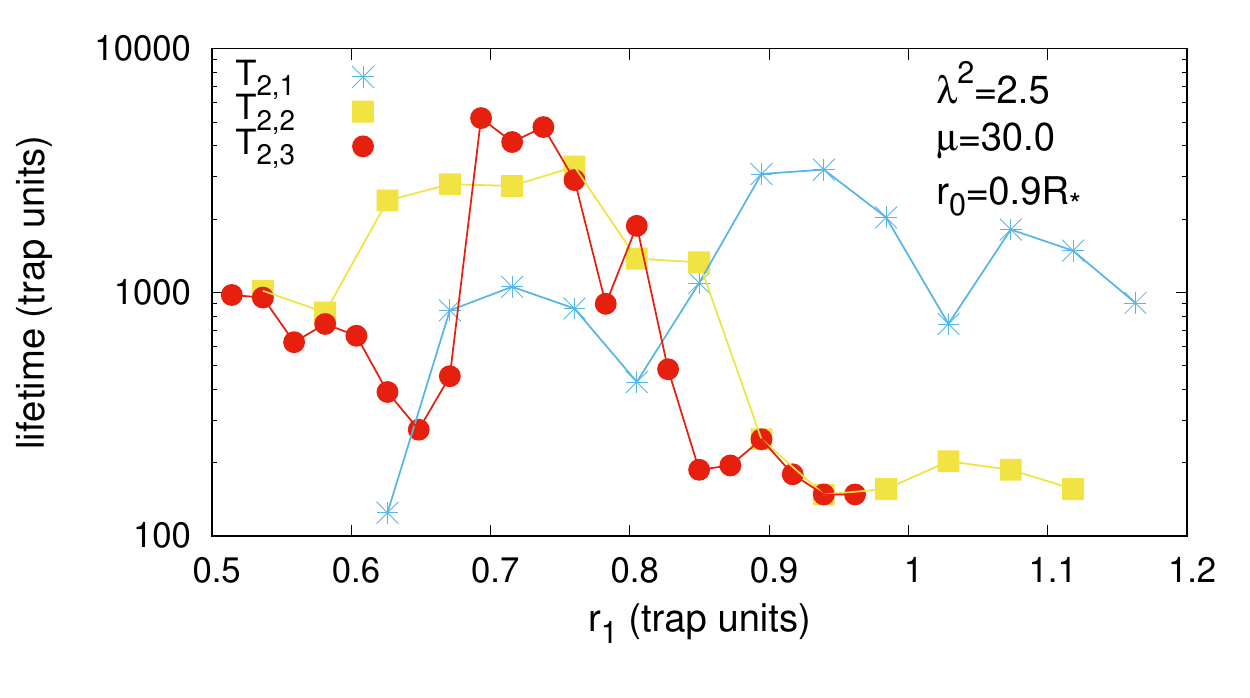, width=88mm}
\epsfig{file=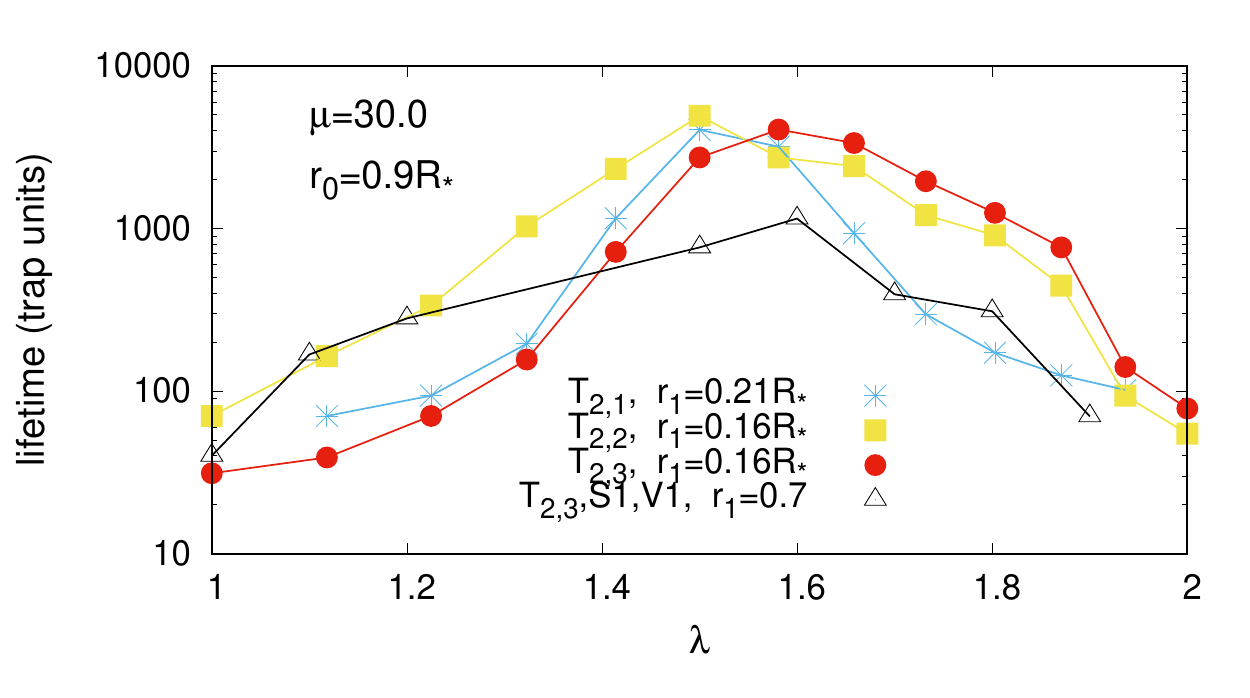,width=88mm}
\end{center}
\caption{Lifetimes over the initial poloidal radius $r_1$  for fixed
anisotropy (top panel) and over the 
anisotropy parameter
$\lambda$ for fixed initial conditions (bottom
panel).}
\label{T_vs} 
\end{figure}

Besides that, to more precisely determine the domains of maximal lifetime within 
the parameter space, we performed several series of simulations with initial states 
prepared using sharp pinning V2 and multiple-mode perturbations S2 for $A_m=0.001R_*$. 
Similarly, we have varied the knot
initial conditions to identify the
lifetime dependence on initial conditions, as well 
as the one on the chemical potential.
In these simulations, the lifetime was measured till the moment of first reconnection.
The results are shown in the top panel of Fig.~\ref{T_vs} 
as lifetime dependencies over the initial poloidal torus radius $r_1$
with a fixed, nearly optimal value of the anisotropy parameter $\lambda$, 
and with a fixed, nearly optimal value of the toroidal radius $r_0$ 
(it should be noted here that optimal $r_0$ has been empirically found
as approximately $0.9R_*$ at moderately large $\mu\sim 30$, 
slightly different from the theoretical limit $1.0 R_*$). In the bottom panel of 
Fig.~\ref{T_vs}, the lifetimes are plotted versus the anisotropy
parameter $\lambda$ for fixed initial $r_1$ and $r_0$. 
For comparison, analogous results for smooth
pinning V1 and single-$m$ perturbation S1 are also shown there.
Indeed distinct parametric intervals can be identified 
where simple vortex knots, unknots, and links 
survive over many hundreds of their revolutions.
Prototypical examples of each class are offered in Fig.~\ref{T_vs}.
For instance, for $1.4<\lambda < 1.8$, we observe the
significant increase of the structure lifetimes (bottom panel).
A similar feature arises for $0.65 < r_1< 0.8$, as a function of
the initial condition parameter $r_1$, for fixed $\lambda$.

Here it should be mentioned that a control simulation with large perturbation
corresponding to $A_m=0.01R_*$ demonstrated decrease of the lifetime in the quasi-stable 
domain by a factor of roughly ten (not shown).
Moderate perturbation with $A_m=0.005R_*$ resulted in roughly two times shorter lifetime
which is still quite long. Thus, even ``less
accurately prepared'' vortex knots are able to 
exist for a long time in suitable parametric regimes,
as revealed by our study.

\begin{figure}
\begin{center}
\epsfig{file=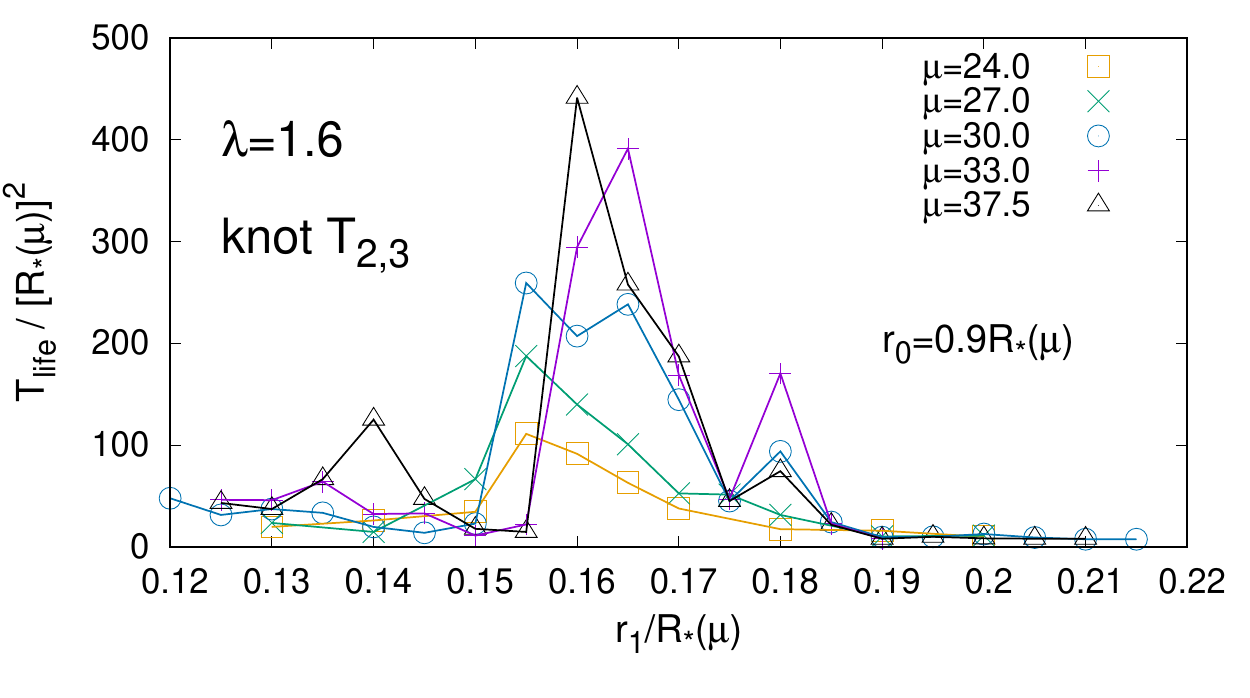, width=88mm}
\end{center}    
\caption{Normalized trefoil lifetimes for different values of the chemical potential. 
The tendency towards larger lifetimes can be observed
for increased values of $\mu$.}
\label{T23life_mu}
\end{figure}

Finally, in Fig.\ref{T23life_mu} we compare trefoil lifetimes as functions of 
the ratio $r_1/R_*$, for different values of chemical potential $\mu$. 
In this case it is convenient to normalize the results to $R_*^2(\mu)$ in order to
separate the overall tendency $T_{\rm life}\sim \mu$. Thus, the
normalized lifetime provides
a general impression (up to a factor of order 1) about the number of knot 
rotations before its destruction. 
We can observe an evident tendency towards an enhanced lifetime
for larger values of the chemical potential $\mu$. This  result is intuitively natural
since a larger $\mu$ implies a weaker coupling of the vortical pattern to sound modes. 

\section{Conclusions}
In this work, we have explored the effects of initial condition
preparation (through variations of the poloidal coordinate $r_1$), trap anisotropy 
(by tuning the confinement ratio $\lambda$) and background density/nonlinearity
(varying the chemical potential $\mu$) to the lifetime of knot
structures in confined atomic Bose-Einstein condensates.
Arguably, our most significant finding is 
that anisotropic traps with (trapping ratios) 
$\lambda\approx 1.6$ can essentially stabilize (i.e., lead
to enhanced lifetimes by over an order of magnitude)
torus vortex knots and links 
in Bose-Einstein condensates with moderate values of the local induction 
parameter $\Lambda=\ln(R_*/\xi_*)\lesssim 3$. 
We similarly identified optimal values of $r_1$ and illustrated
the enhanced lifetime for larger chemical potentials $\mu$.

We observed that the dynamics leading to the eventual demise of the most-long-lived
knots and links involves the sound generation by the rotating vortex structures. 
This process gradually increases the parameter $r_1$ ``pushing it out'' of a quasi-stable 
interval. After that 
Kelvin waves are produced progressively distorting 
the knots/links and ultimately leading to their destruction (via either untying or
leaving the Thomas-Fermi region).
For comparison, recent results based on the Biot-Savart approximation 
indicate that for vortex knots and links of the same relative sizes in spatially 
uniform condensates the mechanism of knot destruction is 
an intrinsic linear instability without stable zones \cite{R2018-2,R2018-3}. 

These results offer, in our view, a systematic understanding
of the viability of observation of torus knots in confined
atomic condensates. They show how initial condition, trapping and nonlinear
features of the underlying problem may enhance the relevant
lifetimes rendering them accessible in current state-of-the-art
experiments. Moreover, they offer some intuition of the relevant
optimality of slightly oblate condensates in connection with
corresponding results for vortex rings. Among the important
open tasks still remaining are of course the experimental realization
of such structures via phase (or perhaps density) engineering, but
also a realization of these knots as exact solutions of the
system from a computation/numerical perspective. In particular,
their rotation suggests that they may be exact periodic orbits
of the system, hence computationally demanding periodic orbit 
identification tools may be used to find such exact solutions
and to assess their stability via, e.g., Floquet theory.
Relevant possibilities will be explored in future studies.

{\it Acknowledgements.} 
This material is based upon work supported by 
the U.S.\ National Science Foundation under Grant No.\ PHY-1602994 
and Grant No.\ DMS-1809074 (P.G.K.). The work of V.P.R. was supported by
the R.F. program No. 0033-2019-0003. 
C.T. ackknowledges funding from the Los Alamos National Laboratory, which is operated by Triad National Security, LLC for the Department of Energy under Contract No.89233218CNA000001.


\begin{thebibliography}{99}

\bibitem{Saffman} P. G. Saffman, {\it Vortex Dynamics} (Cambridge University Press, Cambridge, 1992).

\bibitem{Pismen} L. M. Pismen, {\it Vortices in Nonlinear Fields} (Clarendon, Oxford, 1999).

\bibitem{Donnelly} R. J. Donnelly,  {\it Quantized Vortices in Helium II}
(Cambridge University Press, Cambridge, 1991).

\bibitem{pita1} C. J. Pethick and H. Smith,
{\it Bose-Einstein condensation in dilute gases}, Cambridge University
Press (Cambridge, 2002).

\bibitem{pita2} L. P. Pitaevskii and S. Stringari,
{\it Bose-Einstein Condensation}, Oxford University Press (Oxford, 2003).

\bibitem{ourbook} P. G. Kevrekidis, D. J. Frantzeskakis, and R. Carretero-Gonz{\'a}lez,
{\it The defocusing nonlinear Schr{\"o}dinger equation: from dark solitons
and vortices to vortex rings} (SIAM, Philadelphia, 2015).

\bibitem{F2009}  A. L. Fetter, Rev. Mod. Phys. {\bf 81}, 647 (2009).

\bibitem{FS2001} A. L. Fetter and A. A. Svidzinsky,
J. Phys.: Condens. Matter {\bf 13}, R135 (2001).

\bibitem{PGK2004} P. G. Kevrekidis, R. Carretero-Gonz{\'a}lez,
D. J. Frantzeskakis, and I. G. Kevrekidis,
Mod. Phys. Lett. B {\bf 18}, 1481 (2004).

\bibitem{Kom2007} S. Komineas, Eur. Phys. J. Sp. Top. {\bf 147},
133 (2007).

\bibitem{ParkerBar} C. F. Barenghi and N. G. Parker,
{\it A primer in quantum fluids}, Springer-Verlag (Berlin, 2016).

\bibitem{andAJP} A. C. White, B. P. Anderson, and V. S. Bagnato,
Proc. Nat. Acad. Sci. {\bf 111}, 4719 (2014).

\bibitem{SF2000} A. A. Svidzinsky and A. L. Fetter, Phys. Rev. A {\bf 62}, 063617 (2000).

\bibitem{R2001} V. P. Ruban, Phys. Rev. E {\bf 64}, 036305 (2001).

\bibitem{AR2001} A. Aftalion and T. Riviere, Phys. Rev. A {\bf 64}, 043611 (2001).

\bibitem{GP2001} J. Garcia-Ripoll and V. Perez-Garcia, Phys. Rev. A {\bf 64}, 053611 (2001).

\bibitem{AR2002} A. Aftalion and R. L. Jerrard, Phys. Rev. A {\bf 66}, 023611 (2002).

\bibitem{RBD2002} P. Rosenbusch, V. Bretin, and J. Dalibard, Phys. Rev. Lett. {\bf 89}, 200403 (2002).

\bibitem{AD2003} A. Aftalion and I. Danaila, Phys. Rev. A {\bf 68}, 023603 (2003).

\bibitem{AD2004} A. Aftalion and I. Danaila, Phys. Rev. A {\bf 69}, 033608 (2004).

\bibitem{D2005} I. Danaila, Phys. Rev. A {\bf 72}, 013605 (2005).

\bibitem{Kelvin_waves} A. Fetter, Phys. Rev. A {\bf 69}, 043617 (2004).

\bibitem{ring_instability} T.-L. Horng, S.-C. Gou, and T.-C. Lin,
Phys. Rev. A {\bf 74}, 041603 (2006).

\bibitem{v-2015} S. Serafini, M. Barbiero, M. Debortoli, S. Donadello, F. Larcher, F. Dalfovo, 
G. Lamporesi, and G. Ferrari, Phys. Rev. Lett. {\bf 115}, 170402 (2015).

\bibitem{BWTCFCK2015} R. N. Bisset, W. Wang, C. Ticknor, R. Carretero-Gonzalez, 
D. J. Frantzeskakis, L. A. Collins, and P. G. Kevrekidis,
Phys. Rev. A {\bf 92}, 063611 (2015). 

\bibitem{R2017-2} V. P. Ruban, JETP {\bf 124}, 932 (2017).

\bibitem{R2017-3} V. P. Ruban, JETP Lett. {\bf 106}, 223 (2017).

\bibitem{reconn-2017}
S. Serafini, L. Galantucci, E. Iseni, T. Bienaime, R. N. Bisset, C. F. Barenghi, F. Dalfovo, 
G. Lamporesi, and G. Ferrari, Phys. Rev. X {\bf 7}, 021031 (2017).

\bibitem{top-2017} R. N. Bisset, S. Serafini, E. Iseni, M. Barbiero, T. Bienaime, G. Lamporesi, 
G. Ferrari, and F. Dalfovo, Phys. Rev. A {\bf 96}, 053605 (2017).

\bibitem{WBTCFCK2017} W. Wang, R. N. Bisset, C. Ticknor, R. Carretero-Gonzalez, 
D. J. Frantzeskakis, L. A. Collins, and P. G. Kevrekidis, Phys. Rev. A {\bf 95}, 043638 (2017). 

\bibitem{TWK2018} C. Ticknor, W. Wang, and P. G. Kevrekidis, Phys. Rev. A {\bf 98}, 033609 (2018). 

\bibitem{promentbar} D. H. Wacks, A. W. Baggaley, and C. F. Barenghi,
Phys. Fluids {\bf 26}, 027102 (2014).

\bibitem{talley} R. M. Caplan, J. D. Talley, R. Carretero-Gonz{\'a}lez,
and P. G. Kevrekidis, Phys. Fluids {\bf 26}, 097101 (2014).

\bibitem{R2018PoF} V. P. Ruban, Phys. Fluids {\bf 30}, 084104 (2018).

\bibitem{RSB999} R. L. Ricca, D. C. Samuels, and C. F. Barenghi, J. Fluid Mech. {\bf 391}, 29 (1999).

\bibitem{MABR2010} F. Maggioni, S. Alamri, C. F. Barenghi, and R. L. Ricca, 
Phys. Rev. E {\bf 82}, 026309 (2010).

\bibitem{POB2012} D. Proment, M. Onorato, and C. F. Barenghi, Phys. Rev. E {\bf 85}, 036306 (2012).

\bibitem{KI2013} D. Kleckner and W. T. M. Irvine, Nature Physics {\bf 9}, 253 (2013). 

\bibitem{POB2014} D. Proment, M. Onorato, and C. F. Barenghi, 
J. Phys.: Conf. Ser. {\bf 544}, 012022, (2014).

\bibitem{LMB016} P. Clark di Leoni, P. D. Mininni, and M. E. Brachet, 
Phys. Rev. A {\bf 94}, 043605 (2016).

\bibitem{KKI2016} D. Kleckner, L. H. Kauffman, and W. T. M. Irvine, Nature  Physics {\bf 12}, 650 (2016).

\bibitem{R2018-2} V. P. Ruban, JETP Lett. {\bf 107}, 307 (2018).

\bibitem{R2018-3} V. P. Ruban, JETP {\bf 127}, 581 (2018).

\bibitem{hall1} D. S. Hall, M. W. Ray, K. Tiurev, E. Ruokokoski,
and A.H. Gheorghe, M. M{\"o}tt{\"o}nen, Nat. Phys. {\bf 12}, 478 (2016).

\bibitem{hall2} W. Lee,
A. H. Gheorghe, K. Tiurev, T. Ollikainen, M. M{\"o}tt{\"o}nen, and D.S. Hall,
Science Advances {\bf 4}, 3820 (2018).

\bibitem{R2018-1} V. P. Ruban, JETP  {\bf 126}, 397 (2018).

\bibitem{R2018-4} V. P. Ruban, JETP Lett. {\bf 108}, 605 (2018).

\bibitem{note} In traps the length scale is $l_r=\sqrt{\hbar/m \omega_r}$ where $\omega_r$ 
is trapping frequency, $m$ is mass. The energy scale is $\hbar\omega_r$ and time scale is $1/\omega_r$.

\bibitem{hopflink} https://en.wikipedia.org/wiki/Link\_(knot\_theory)

\bibitem{FBD2010} C. J. Foster, P. B. Blakie, and M. J. Davis,
Phys. Rev. A {\bf 81}, 023623 (2010).

\end{thebibliography}
\end{document}